\documentclass[conference]{IEEEtran}

\IEEEoverridecommandlockouts                              

\usepackage{graphicx} 
\usepackage{bm}
\usepackage{amsmath}
\usepackage{algpseudocode}
\usepackage{algorithm}
\usepackage{amssymb}
\usepackage{optidef}
\usepackage{xcolor}
\usepackage{comment}
\usepackage{caption}
\usepackage[a4paper, total={8.5in, 11in}, portrait, margin=1in]{geometry}
\def\BibTeX{{\rm B\kern-.05em{\sc i\kern-.025em b}\kern-.08em
    T\kern-.1667em\lower.7ex\hbox{E}\kern-.125emX}}

\title{
Optimizing Resource Allocation and Scheduling towards FRMCS and GSM-R networks coexistence in Railway Systems
}

\begin{document}

\author{
\IEEEauthorblockN{Mohamed Aziz Aboud\IEEEauthorrefmark{1}\IEEEauthorrefmark{3}, 
Nawel Zangar\IEEEauthorrefmark{3}, Rami Langar\IEEEauthorrefmark{1}\IEEEauthorrefmark{3}, Marion Berbineau\IEEEauthorrefmark{3}, and Jerome Madec\IEEEauthorrefmark{4} }

\IEEEauthorblockA{\IEEEauthorrefmark{1} Software and IT Engineering Department, Ecole de Technologie Supérieure (ÉTS), Montréal, QC H3C 1K3, Canada}

\IEEEauthorblockA{\IEEEauthorrefmark{3} LIGM-CNRS UMR 8049, University Gustave Eiffel, F-77420 Marne$-$la-Vallée, France}

\IEEEauthorblockA{\IEEEauthorrefmark{4} SNCF-Réseaux, Direction Télécom Unifiée, 93574 La Plaine-St-Denis, France}

\IEEEauthorblockA{
E-mails: {mohamed-aziz.aboud.1@ens.etsmtl.ca; rami.langar@etsmtl.ca;}}
\IEEEauthorblockA{
{\{nawel.zangar, marion.berbineau\}@univ-eiffel.fr; jerome.madec@reseau.sncf.fr}}
\vspace{-5mm}
}

\maketitle

\thispagestyle{empty}
\pagestyle{empty}

\begin{abstract}

The actual railway communication system used in Europe for high-speed trains (HST) is called the GSM-R system, which is a communication system based on 2G infrastructure. This system is meant to be replaced by a new system based on 5G NR infrastructure called the Future Railway Mobile Communication System (FRMCS) by 2030. For the next years, both systems will probably coexist in the same frequency band since the migration from GSM-R to FRMCS is planned to be done progressively until the GSM-R system is completely shut down, mainly due to safety and budget constraints.
In this paper, we study the resource allocation for the FRMCS system sharing the same frequency band as the already deployed GSM-R system. We formulate the resource allocation problem as an integer linear problem (ILP), known to be NP-hard.To solve it in a reasonable time, we propose a scheduling algorithm, called Intelligent Traffic Scheduling Preemptor (ITSP), that allocates resources for the different FRMCS traffic types considered (critical traffic and performance traffic) in the same frequency band with the GSM-R system. Our algorithm is channel quality Indicator (CQI) aware and uses the preemption mechanism in 5G NR standards to optimize the resource allocation for the FRMCS system without impacting the actual GSM-R resource allocation in the context of the white space concept.

\end{abstract}

\vspace{-3mm}
\section{Introduction}

With the rapid development of 5G New Radio (NR) and beyond communication systems, a new railway communication system for High-Speed Trains is also being developed. The new system called the Future Railway Mobile Communication System (FRMCS) is a dedicated system for railway based on 5G NR, particularly in terms of the core network infrastructure. FRMCS is meant to replace progressively the actual railway communication system, GSM-R. The migration between the two systems is planned to be done progressively mainly due to safety and budget constraints. Thus, for the next years until at least 2030, where it is supposed to start the progressive shut down of GSM-R, both systems will coexist, sharing a part of the bandwidth dedicated to them.

In order to make this migration from GSM-R to FRMCS in the best conditions there are two main principles that need to be respected.

The first principle for this spectrum sharing is to preserve, as most as possible, the GSM-R for both uplink (UL) and downlink (DL). The second principle consists in accepting some form of GSM-R to 5G interference impacting the 5G performances, but reducing any form of 5G to GSM-R interference. 5G is strong enough in term of channel coding mechanisms (LDPC, HARQ, repetition strategies) to support this kind of additional interference.

A key and challenging issue tackled in this paper is the scheduling algorithm responsible for allocating resource blocks to both FRMCS and GSM-R users, while respecting users' QoS.

The existing scheduling techniques are not adapted to the actual context of different coexisting technologies sharing the same spectrum. Therefore, we propose in this paper a novel scheduling algorithm to provide an adequate solution to this problem using the channel quality, the preemption technique and taking into consideration the FRMCS resource blocks colliding with GSM-R carriers.

The main contributions of this paper can be summarized as follows:
\begin{itemize}
    \item We define the GSM-R, FRMCS coexisting and spectrum sharing problem using the white space concept, presenting the available shared frequency resources and the colliding resource blocks.

    \item We formulate the resource allocation problem as an Integer Linear Problem (ILP), aiming to maximize the throughput of performance traffic applications, while guaranteeing the transmission of both low latency critical and GSM-R traffics.

    \item We propose a scheduling algorithm to solve the previously mentioned problem with low time complexity, namely the Intelligent Traffic Scheduling Preemptor (ITSP) algorithm.

    \item Through simulations, we demonstrate the efficiency of our proposed solution and its suitability for our railway communication system in the 2G-5G migration context.
\end{itemize}

The rest of the paper is organized as follows. Section II describes the background and related work introducing the white space concept and the existing schedulers in 5G NR. Section III presents the system model and our problem formulation as an ILP. Section IV presents the proposed scheduling algorithm and explains how it works. The algorithm performance evaluation and the simulation results are reported in Section V. Finally, section VI concludes the paper by summarizing the key findings and contributions.

\section{Background and Related work}
\subsection{Background: White space concept and FRMCS deployment specifications}

The "white space" concept is used for shared spectrum access, where FRMCS is deployed over existing GSM-R frequencies. In current railway deployments, GSM-R operates on uplink (UL) frequencies of [876.0 MHz, 880.0 MHz] and downlink (DL) frequencies of [921.0 MHz, 925.0 MHz]. Meanwhile FRMCS utilizes [874.4 MHz, 879.4 MHz] for UL and [919.4 MHz, 924.4 MHz] for DL. Notably, there is an overlap in frequency ranges from 876 to 879.4 MHz for UL and from 921 to 924.4 MHz for DL ; outside this range, there are no conflicts. 

\begin{figure}[t]%
\includegraphics[width=0.48\textwidth]{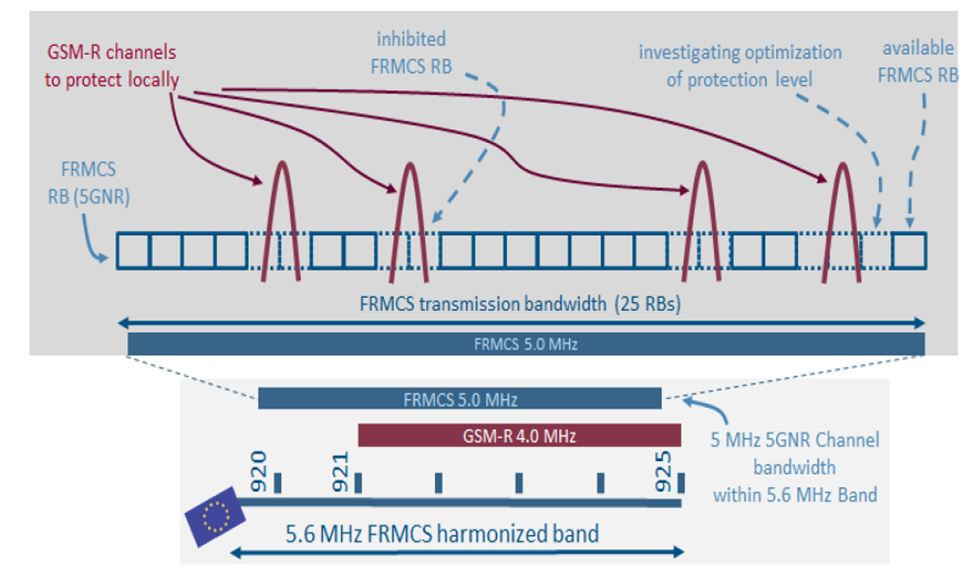}
{\caption{FRMCS and GSM-R co-existence with white space concept \cite{bib2}}
\label{fig2}}
\vspace{-4mm}
\end{figure}

Thanks to the flexible 5G OFDM structure, this 2G-5G coexistence is possible. Figure \ref{fig2} illustrates how the FRMCS system is implemented in the same frequency as the existing GSM-R system taking into consideration the white spaces corresponding to the gaps between deployed GSM-R channels.


The most popular existing schedulers, such as round robin (RR) \cite{bib3} and proportional fair (PF)\cite{bib4} are not adequate for such 2G-5G coexistence system, since they aim to provide fairness between different users in term of throughput, which will not be suitable for our case. 

In the next subsection, we will review most relevant works related to the scheduling techniques in 5G NR.



\subsection{Scheduling in 5G NR}

Resource allocation in 5G NR aims to maximize user QoE by efficiently distributing limited radio resources. In the OFDM resource grid, resources are allocated across time (based on transmission duration) and frequency (carrier frequency and bandwidth). The gNB scheduler manages this process to optimize resource allocation for user equipment (UEs).
To improve data rates, CQI-aware schedulers consider channel quality indicators (CQI) when assigning PRBs. The authors in \cite{bib5} propose a CQI-aware scheduler that prioritizes high-CQI code blocks for users in each network slice based on SLA requirements. This approach significantly enhances data rates compared to non-CQI-aware schedulers.
For low-latency, high-reliability critical traffic, 5G NR uses the puncturing/preemption mechanism \cite{bib13}. As described in \cite{bib6}, resources are first allocated to eMBB UEs, with some mini slots of these PRBs dynamically replaced by URLLC data if needed to meet latency requirements while minimizing eMBB throughput loss.

Inspired from the two latter concepts, we propose in this paper a new CQI-aware scheduling algorithm for joint FRMCS and GSM-R systems that uses the preemption method to prioritize the FRMCS critical-type traffic over the performance-type one without impacting GSM-R traffic, as will be explained in the next section. By doing so, we increase the reliability of critical-type transmissions.





\section{System model and Problem Formulation} \label{model_problem}
\subsection{System Model}\label{System Model}

To address resource block allocation for FRMCS, without impacting the GSM-R traffic, we incorporate channel quality assessments based on relevant propagation and path loss models for the high-speed rail (HSR) context. We introduce two path loss models commonly used on railway scenarios : The Urban Macro model (UMa)\cite{bib7} and the Rural Macro model (RMa)\cite{bib8}.

\begin{figure}[t]%
\includegraphics[width=0.48\textwidth]{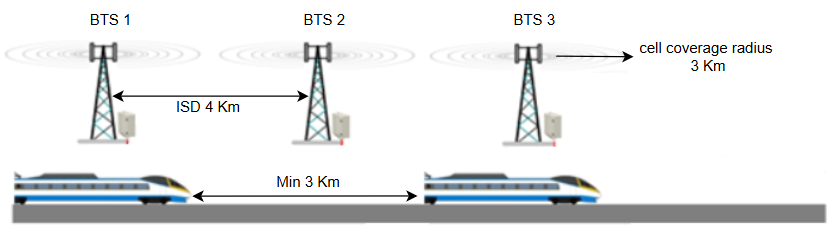}
{\caption{Railway infrastructure}
\label{fig4}}
\vspace{-4mm}
\end{figure}

The railway infrastructure is illustrated in Figure \ref{fig4}. Base stations (gNB) are spaced approximately by 4 km apart in urban areas and 8 km in rural areas, each covering about a 3 km radius along the railway. As Trains, communicate with gNB while in motion, path loss is used to calculate the corresponding Signal-to-Noise-Ratio (SNR) to estimate the channel quality. This information is then used to determine the appropriate modulation for data transmission.

For the network traffic model, we consider two types of FRMCS traffic commonly used in railway scenarios: critical traffic and performance traffic. Critical traffic is mainly divided into data traffic, such as signaling for the European Train Control System (ETCS), and voice traffic, mainly for emergency calls. This type of traffic has a low latency constraint ($\leq$ 100 ms) but does not require a high data rate (around 100 kbps).
The performance traffic is less prioritized than the critical traffic, it does not require a low latency, but it has a high data rate constraint (up to 10 Mbps for applications like standard data communications).
In the following, we formulate our scheduling problem as an ILP.

\subsection{Problem Formulation}

\begin{figure}[t]
\centerline{\includegraphics[width=0.48\textwidth]{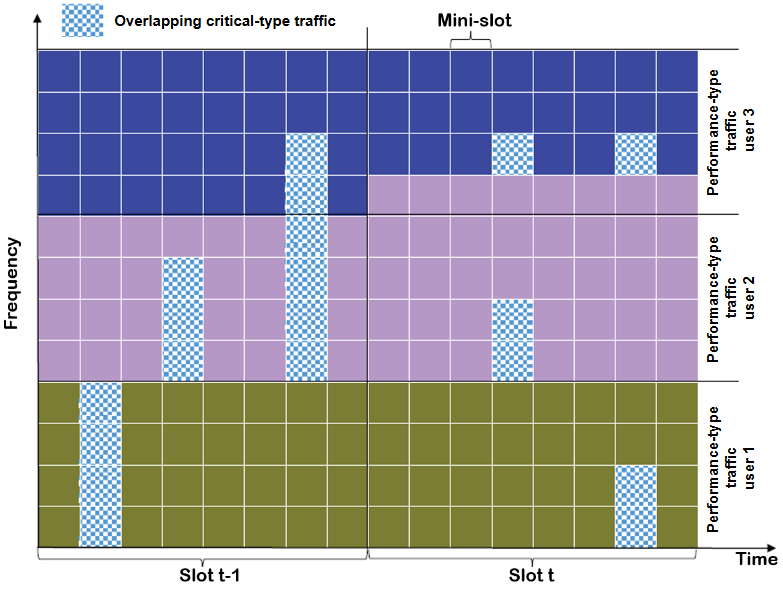}}
\caption{Multiplexing between performance-type and critical-type traffics}
\vspace{-4mm}
\label{fig12}
\end{figure}

For the sake of comparison with common resource and scheduling approaches, let us first show how our problem with and without the preemption method could be formulated as an ILP optimization problem.

As stated earlier, since we are considering two different applications profiles in railway systems, we first define two set of users $\mathcal{U}_p^{j}$ and $\mathcal{U}_c^{j}$, all attached to the gNB $j$ and initiating performance-type traffic and critical-type traffic, respectively.

We assume that Physical Resource Blocks (PRBs) are allocated for performance-type traffic on a time-slot basis, while they are managed for critical-type traffic on a mini-slot basis \cite{bib12}, as shown in Fig. \ref{fig12}. We hence consider a slotted, synchronized, and periodic FRMCS network with time period \(T\). This time period is divided in \(|T|\) time slots; each time slot \(t\) is divided into \(|M|\) mini-slots of the same length in order to provide low latency services, such as critical-type traffic. 

We consider \(|T|\) time slots to serve performance-type traffic users, represented by \(T = \{1,2,\) ... \( ,|T|\}\). We assume that critical-type traffic arrives at a gNB $j$ in any mini-slot \(m\) of time slot \(t\) following a random process (e.g. Poisson process), and its payload size is varying from 70 to 120 Bytes, as reported in \cite{bib10}. We also assume that GSM-R traffic arrives at the gNB $j$ following a Poisson process and uses a predefined set of PRBs.

We analyze the resource allocation through two binary variables $x_{i,k}^{t}$ and $y_{i,k}^{m,t}$ . The first indicates whether the PRB $k \in \{1, ..., K\}$ is allocated to user $i \in \mathcal{U}_p^{j}$ at time slot $t$. The latter indicates whether the PRB $k \in \{1, ..., K\}$ is allocated to user $i \in \mathcal{U}_c^{j}$ at mini-slot $m$ of time slot $t$, where $K$ is the total number of available $PRBs$ in the network. 

Also, let $E_c$ be the set of potential colliding PRBs that are shared between FRMCS and GSM-R systems. To represent the PRBs that are potentially colliding with GSM-R, we define another binary variable $a_k$ to indicate if a PRB $k$ is potentially in collision with a GSM-R channel.
\vspace{-3mm}\\
\begin{equation}
  a_{k} =
    \begin{cases}
      1 & \text{if $PRB~k \in E_c ;  \forall k \in \{1, ..,K\}$ }\\
      0 & \text{otherwise}
    \end{cases}       
\end{equation}
\vspace{-4mm}\\
To represent the PRBs that are currently being used by the GSM-R traffic during the scheduling period $T$, we define another binary variable $b_k$ as follows:
\vspace{-3mm}\\
\begin{equation}
  b_{k} =
    \begin{cases}
      1 & \text{if $PRB~k$ is used by GSM-R; $\forall k \in E_c$ }\\
      0 & \text{otherwise}
    \end{cases}       
\end{equation}
\vspace{-6mm}\\
Given the users' position, the maximum achievable throughput $\gamma_{i,j}^k$ of the user $i$ attached to the gNB $j$ on the PRB $k$ can be expressed as follows:
\vspace{-4mm}\\
\begin{equation}
\gamma_{i,j}^k= B_k \log_{2} (1 + \Gamma_{i,j}^k)
\end{equation}
where $B_k$ is the PRB bandwidth (corresponding to 180 kHz in the FRMCS system), and $\Gamma_{i,j}^k$ is the Signal-to-Noise-Ratio (SNR) between the user $i$ and its attached gNB $j$ on the PRB $k$ (which is computed as explained in subsection \ref{System Model}).

To enable the preemption method, we introduce the binary variable \(z_{j,k}^{m,t}\) to indicate if the gNB \(j\) is using the PRB \(k\) during the mini-slot \(m\) of a slot \(t\) to transmit only the performance-type traffic.
\vspace{-3mm}\\
\begin{equation}
    z_{j,k}^{m,t}= \begin{cases}
    1 & \textit{if }  x_{i,k}^{t} - y_{i',k}^{m,t} = 1; \forall i \in \mathcal{U}_p^{j}, \forall i' \in \mathcal{U}_c^{j} \\
    0 & \text{otherwise} 
    \end{cases}
    \label{eq13}
\end{equation}
To know if a gNB \(j\) transmits performance-type traffic during the mini-slot \(m\) of slot \(t\), we use the binary variable \(z_{j}^{m,t}\):
\vspace{-3mm}\\
\begin{equation}
    z_{j}^{m,t}= \begin{cases}
    0 & \textit{if }  \displaystyle \sum_{k=1}^{K} z_{j,k}^{m,t} = 0; \forall j, m, t \\
    1 & \text{otherwise}
    \end{cases}
    \label{eq14}
\end{equation}
\vspace{-4mm}\\
The throughput of the performance-type traffic is given by the ratio of the traffic successfully routed to users $i \in \mathcal{U}_p^{j}$ and the number of required time-slots. Therefore, maximizing throughput can be achieved by minimizing the total number of time slots used during the period \(T\) to transmit the performance-type traffic.

Hence, our resource allocation problem in the context of coexistence between FRMCS and GSM-R systems and with enabling the preemption method at a given gNB $j$ can be formulated as follows: 
\begin{footnotesize}
\begin{maxi*} |s| 
		{\hspace{-2mm} x, y, z} {\hspace{-3mm} \frac{\displaystyle \sum_{t=1}^{|T|}\sum_{k=1}^{K}\sum_{i=1}^{|\mathcal{U}_p^{j}|}x_{i,k}^{t}}{|T|\times K \times |\mathcal{U}_p^{j}|} - \frac{\displaystyle \sum_{t=1}^{|T|}\sum_{m=1}^{|M|}\sum_{k=1}^{K}\sum_{i=1}^{|\mathcal{U}_c^{j}|}y_{i,k}^{m,t}}{|T|\times |M| \times K \times |\mathcal{U}_c^{j}|}
         - \frac{\displaystyle \sum_{t=1}^{|T|}\sum_{m=1}^{|M|} z_{j}^{m,t}}{|T|\times|M|}} 
        {}{} \tag{10}
		\addConstraint {\sum_{i=1}^{|\mathcal{U}_p^{j}|} x_{i,k}^{t} \leq 1, \hspace{2mm} \forall t, \hspace{2mm} \forall k \in \{1, ..., K\}} \label{c1} \tag{a}
        \addConstraint {\sum_{i=1}^{|\mathcal{U}_c^{j}|} y_{i,k}^{m, t}  \leq 1, \hspace{2mm} \forall m, \hspace{2mm} \forall t, \hspace{2mm} \forall k \in \{1, ..., K\}} \label{c2} \tag{b}
        \addConstraint {x_{i,k}^{t} \leq (1 - b_k), \hspace{2mm} \forall k \in \{1, ..., K\}, \forall i \in \mathcal{U}_p^{j}} \label{c4} \tag{c}
        \addConstraint {y_{i,k}^{m,t} \leq (1 - a_k), \hspace{2mm} \forall k \in \{1, ..., K\}, \forall i \in \mathcal{U}_c^{j}} \label{c5} \tag{d}
        \addConstraint {\sum_{t=1}^{|T|} \sum_{k=1}^{K} \frac{\displaystyle \sum_{m=1}^{|M|} z_{j,k}^{m,t}}{|M|} \times
        x_{i,k}^{t} \times \gamma_{i,j}^k \leq TH_i^{perf} \hspace{2mm} \forall i \in \mathcal{U}_p^{j}} \label{c6} \tag{e}
        \addConstraint {\sum_{t=1}^{|T|} \sum_{m=1}^{|M|} \sum_{k=1}^{K} y_{i,k}^{m,t} \times \gamma_{i,j}^k \geq TH_i^{critic} \hspace{2mm} \forall i \in \mathcal{U}_c^{j}}  \label{c7} \tag{f}
        \addConstraint \sum_{t=1}^{|T|}\sum_{m=1}^{|M|} y_{i,k}^{m,t} \leq D_i^{critic} \hspace{2mm} \forall k, \hspace{2mm} \forall i \in \mathcal{U}_c^{j}   \label{c8} \tag{g}
        {}
\end{maxi*}
\end{footnotesize}%


The objective function in (10) contains three terms. The first term (respectively, the second term) expresses the total number of allocated PRBs for the performance-type traffic (respectively, the critical-type traffic) during the scheduling period $T$. While the third term expresses the total number of time slots used during the period $T$ to transmit the performance-type traffic. Our aim is to maximize the first term and at the same time minimize the second and third terms, while satisfying users' QoS. It is worth noting that all these terms have been normalized.

Constraints (\ref{c1}) and (\ref{c2}) ensure that two users initiating the same type of applications cannot use the same PRB. Constraint (\ref{c4}) indicates that performance-type traffic users can use the colliding PRBs that are actually not being used by the GSM-R traffic, while constraint (\ref{c5}) indicates that critical-type traffic users must use the non-colliding PRBs. Constraint (\ref{c6}) indicates that a performance-type traffic user can not obtain more than what he demands. In this constraint, we need to consider only the fraction of the time slot that is actually being used by this type of traffic, since it can be preempted by an incoming critical-type traffic using any mini-slot. Finally, constraints \eqref{c7} and \eqref{c8} reflect the QoS requirements for critical-type traffic in terms of minimum required throughput and maximum scheduling delay, respectively.
Note that, $TH_i^{perf}$ and $TH_i^{critic}$ refer to the required throughput per user $i$ for the performance-type traffic and critical-type traffic, respectively. Whereas $D_i^{critic}$ refers to the required scheduling delay for critical railway applications, such as emergency calls and signaling data exchange for European Train Control System (ETCS). These parameters are given by the railways system's administrator.

It is worth noting that the constraint (e) is not linear. To linearize it, we introduce a new binary variable $f_{i,k}^{m,t}$ which represents the value of $z_{j,k}^{m,t} \times x_{i,k}^{t}$. The constraint (e) will be thus replaced by the following four constraints:
\begin{equation}\tag{e-1}
\sum_{t=1}^{|T|} \sum_{m=1}^{|M|} \sum_{k=1}^{K}  f_{i,k}^{m,t} \times \gamma_{i,j}^k \leq |M| \times TH_i^{perf} \hspace{2mm} \forall i \in \mathcal{U}_p^{j}
\end{equation}
\begin{equation}\tag{e-2}
    f_{i,k}^{m,t} \leq z_{j,k}^{m,t}
\end{equation}
\begin{equation}\tag{e-3}
    f_{i,k}^{m,t} \leq x_{i,k}^{t}
\end{equation}
\begin{equation}\tag{e-4}
    f_{i,k}^{m,t} \geq z_{j,k}^{m,t} + x_{i,k}^{t} - 1
\end{equation}

In addition, to relax the preemption method, the following constraint needs to be added:
\begin{equation} \tag{h} \label{c9}
x_{i,k}^{t} + y_{i,k}^{m,t} \leq 1, \hspace{2mm} \forall k, m, t, ~and~ \forall i \in \mathcal{U}_p^{j} \cup \mathcal{U}_c^{j} 
\end{equation}
Where constraint \eqref{c9} allows orthogonal resource allocation for the critical-type traffic with respect to the performance-type one. 

In the next section, we introduce our Intelligent Traffic Scheduling Preemptor (ITSP) algorithm to solve the above formulated ILP problem. Our algorithm uses the channel quality indicator (CQI) and the preemption technique to provide a low complexity and efficient solution of our resource allocation problem taking into consideration the existence of GSM-R traffic and colliding PRBs.

\section{Proposed Resource and Scheduling Algorithm}

Given the NP-hardness of problem (10) and considering its high resolution time as will be shown in Table \ref{optimal table}, we propose the following heuristic algorithm to provide a near optimal solution with a low time complexity. 
Our proposed algorithm, called Intelligent Traffic Scheduling Preemptor (ITSP), is channel quality-aware that prioritizes critical traffic over performance traffic using the preemption technique, while managing resource allocation for FRMCS users within the bandwidth already occupied by the GSM-R system. We hence consider two types of PRBs: i) Collision-Free PRBs, which are exclusively used by FRMCS users, and ii) Colliding PRBs, which are shared with active GSM-R channels. In this case, GSM-R users have higher priority to use them.


Our ITSP algorithm, presented in Algorithm \ref{alg:cap}, operates in three steps, as follows:

Step I (Lines 1 - 9): Identifying Colliding PRBs: For this step, we use a channel sounding technique \cite{bib14} to detect currently occupied GSM-R channels. Doing so, colliding PRBs will be tagged to ensure they remain unused by FRMCS users. Due to space limitation, this technique is not detailed here. 

Step II (Lines 10 - 21): Resource allocation for the performance-type traffic: For the remaining untagged PRBs, we assess performance traffic for FRMCS users. To do so, we first calculate path loss based on the distance from users to the base station (gNB) and determine the signal-to-noise ratio (SNR) and modulation coding scheme (MCS). Then, we allocate PRBs providing the best channel quality indicator (CQI) to corresponding users starting with the user with the best CQI.

Step III (Lines 22 - 42): Resource allocation for critical-type traffic: This step addresses two types of critical traffic: ETCS data and voice traffic, both with a short delay constraint of about 100 ms. To do so, we first allocate available mini-slots (empty mini-slots or mini-slots from allocated PRBs in step II without exceeding the corresponding resource block allowance rate) from collision-free PRBs to FRMCS users with critical traffic, prioritizing control traffic over voice traffic when delays are equal. If there is still unallocated critical traffic with zero delay, we use the preemption technique to replace the mini-slots of PRBs already allocated to performance-type traffic (in step II) with this critical traffic, starting from the PRB with the lowest MCS in order to minimize the impact on the overall network performance.

\begin{algorithm}
\caption{Intelligent Traffic Scheduling Preemptor (ITSP) algorithm}\label{alg:cap}
\begin{algorithmic}[1]
\While {$gsmrTraffic()$ != null do}
  \State $PRB_i \gets CollidingNotTagged(PRBs) $ ;
  \If{$PRB_i$ is not null}
	\State $PRB_i$ is used for gsm-r traffic 
        \State $Tag(PRB_i)$ ; \Comment{$PRB_i$ is tagged and can not be used for FRMCS users}
  \Else{}
	\State break ;
  \EndIf
  \EndWhile
\State $cqi \gets calculateCQI(UEs)$
\For{each user in sortedUsers} \Comment{users are sorted starting with the highest CQI}
  \State $rbNumber \gets PRBNumberNeeded()$      
  \While {$perfTraffic()$ != null and allocatedRBs $< PRBNumber$}
   	\State $PRB_i$ = getAvailableRB(PRBs) ;
	\If {$PRB_i$ is not null}
		\State $PRB_i$ is allocated for perfTraffic traffic ;
	\Else
		\State break ;
        \EndIf
  \EndWhile
\EndFor  
 \State $S_{outage} \gets \emptyset$
 \State sort($Q_{criticalTraffic}$) \Comment{starting with signaling traffic then voice traffic beginning with lowest delay} 
 \While {$criticalTraffic()$ is not null}
 	\For {t in T do}
 		\For {i in F do}
 			\If {CB($PRB_i$,t) in $S_{outage}$} \Comment{check if the code block (CB) preemption capacity is entirely used}
 				\State $PRB_i \gets pop(Q_{URLLC})$ ;
 				\State break ;
                \EndIf
 			\If {$PA(CB(PRB_i,t)) \geq 1$ do} \Comment{check if the are still resource elements (REs) that can be used  for preemption in the PRB without exceeding the preemption allowance (PA)}
 				\State $PRB_i \gets getNotCollidingRB(cqi, user)$ ;
 				\State $PRB_i \gets pop(QURLLC)$ ;
 			    \State break ;
                \EndIf
            \EndFor
        \EndFor
    \EndWhile
\While{$TopPacketLatency(Q_{URLLC}) == 0$ do}
   \State $PRB_i \gets  NotCollidingLowestCQI(cqi, user)$ ;
   \State $PRB_i \gets pop(Q_{URLLC})$ ;
\EndWhile
\\ \textbf{return} ;
\end{algorithmic}
\end{algorithm}

\section{Performance evaluation}

In this section, we assess the effectiveness of our proposed resource allocation algorithm based on extensive simulations using our own simulator developed in python. First, we describe the simulation environment setup. Then, we analyze the obtained results and discuss the effectiveness of our proposal compared to three baselines:
\begin{itemize}
    \item Optimal approach with Preemption by resolving our above-mentioned ILP formulation using the GUROBI solver and where the preemption technique is enabled. We call this approach Opt. Preempt.
    \item Optimal approach without Preemption by resolving the relaxed ILP formulation using the GUROBI solver too and where the preemption technique is disabled by adding the constraint (h). We call this approach Opt. no Preempt.
    \item Best CQI (BCQI) approach proposed in \cite{bib11} which we adapted to only use collision-free PRBs for FRMCS.
\end{itemize}

\subsection{Simulation setup}

We consider a railway infrastructure with one gNB and multiple trains (users). We consider 17 resource blocks for the FRMCS system (out of 25 resource blocks in total within the considered 5.6 MHz bandwidth, from which we remove 8 PRBs reserved for control signals (PUCCH and PRACH)). Since in FRMCS we have 1 slot per radio sub-frame and each slot contains 14 symbols, we consider $M$ = 7 mini-slots assuming 2 symbols per mini-slot \cite{bib15}. We also consider 10 time slots (T) in our simulations corresponding to the radio frame duration of 10 ms. In addition, we consider performance packet sizes of 200 Bytes and critical packet sizes of 100 Bytes \cite{bib10}, with a delay varying of 5 ms for each packet which corresponds to the required delay for critical railway applications, such as emergency calls (with an average required throughout of 300 Kbps \cite{bib16}) and signaling data exchange for European Train Control System (ETCS). \cite{bib16}). 
In addition, we assume that the desired performance traffic throughput for each user is about 10 Mbps \cite{bib16} which corresponds to the required throughput for standard data communication applications.

We use a Poisson arrival model to describe the different traffic types arrival rates (i.e., performance-type, critical-type, and GSM-R traffics). The transmission power of the train's antenna is 23 dBm. The minimal distance between two trains on the same track is set to 3 km (security distance). Table \ref{table1} summarizes the simulation setup parameters.
\begin{table}[t]
\begin{center}
\begin{tabular}{ |c|c| } 
 \hline
 Duration (T) & 10 ms\\
 \hline
 Total available PRBs & 17 \\
 \hline
 Mini-slots number (M) & 7 \\
 \hline
 Colliding PRBs & 2 - 10 \\ 
 \hline
 Critical traffic packet size & 100 Bytes \\ 
 \hline
 Performance traffic packet size & 200 Bytes \\  
 \hline
 $D_i^{critic}$ & 5 ms \\ 
  \hline
$TH_i^{critic}$ & 300 Kbps \\
  \hline
 $TH_i^{perf}$ & 10 Mbps \\
  \hline
\end{tabular}
\end{center}
\caption{Simulation parameters}
\label{table1}
\end{table}
\begin{table}[t]
\begin{center}
\begin{tabular}{|c@{ }|c|c|c|c| } 
 \hline
 Number of users & 1 & 10 & 100 \\
 \hline
 Opt. Preempt computation time (sec) & 0.10 & 0.24 & 2.10 \\
 \hline
 Opt. no Preempt computation time (sec) & 0.35  & 0.06 & 7.21 \\
 \hline
\end{tabular}
\end{center}
\caption{Optimal solutions resolution time}
\vspace{-3mm}
\label{optimal table}
\end{table}

\subsection{Simulation results}
We present the simulation results corresponding to the different scenarios we simulated. We consider two scenarios: a high load scenario (respectively, a low load scenario) with a high load of critical traffic corresponding to 1 Mbps throughput per user (respectively, a low load of critical traffic corresponding to 300 Kbps throughput per user). In both scenario, the GSM-R traffic is assumed to be constant and equals to 200 Kbps. Table \ref{table2} summarizes the simulation parameters specific to each scenario.

\begin{table}[t]
\begin{center}
\begin{tabular}{ |c|c|c| } 
 \hline
\textbf{Parameters} & \textbf{High load} & \textbf{Low load} \\
 \hline
 Path loss model & \multicolumn{2}{|c|}{RMa} \\ 
 \hline
 Number of users (trains) & \multicolumn{2}{|c|}{2} \\
  \hline
 Colliding PRBs number & \multicolumn{2}{|c|}{2 - 10}\\
 \hline
 $\lambda_1$ (performance traffic) & \multicolumn{2}{|c|}{50} \\
 \hline
 $\lambda_2$ (Critical traffic) & 10 & 3\\ 
 \hline
 $\lambda_3$ (GSM-R traffic) & \multicolumn{2}{|c|}{2} \\
\hline
\end{tabular}
\end{center}
\caption{Simulation parameters of different scenarios}
\label{table2}
\end{table}

We consider the presence of 2 UEs (trains) running at a speed of about 300 km/h, each UE initiates both a performance traffic and a critical traffic flow. We consider 2 UEs, reflecting the case of high speed running trains, which typically do not exceed 2 trains communicating with the same gNB, while considering the minimal distance between trains and the gNB coverage area.
The number of UEs can increase into 50 or 60 when we consider trains 
at a major railway station.

\begin{figure}[t]%
\begin{center}
\includegraphics[width=0.42\textwidth]{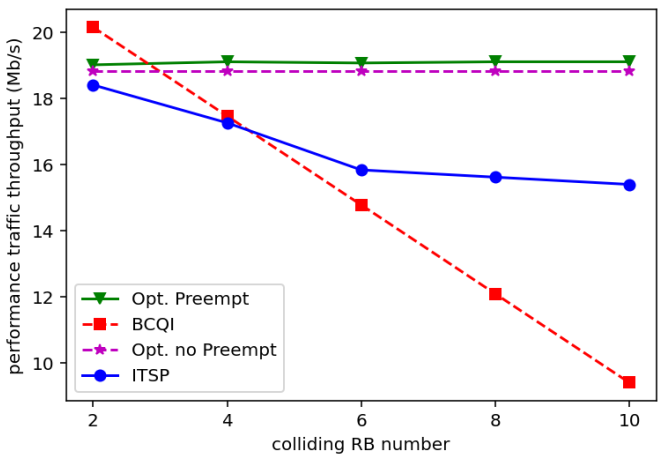} \\
(a) High load scenario ($\lambda_1 = 50, \lambda_2 = 10, D^{critic} = 5$)
\vspace{5mm} \\
\includegraphics[width=0.42\textwidth]{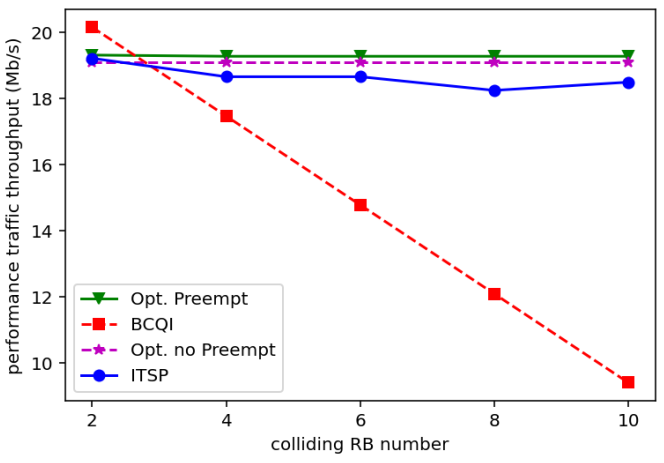} \\
(b) Low load scenario ($\lambda_1 = 50, \lambda_2 = 3, D^{critic} = 5$)
\end{center}
{\caption{Total Network Performance traffic throughput vs. Nbr. of colliding PRBs}
\label{fig15}}
\vspace{-3mm}
\end{figure}

Fig. \ref{fig15} shows the total network performance traffic throughput as a function of colliding PRBs for all approaches. The ITSP algorithm maintains throughput close to optimal solutions, even with more colliding PRBs. In contrast, the BCQI, which doesn’t account for colliding PRBs in FRMCS traffic allocation, sees a performance decline. ITSP, however, allows performance traffic to use potentially colliding PRBs unused by GSM-R, keeping overall network performance acceptable, slightly decreased compared to optimal solutions. 

\begin{figure}[t]%
\begin{center}
\includegraphics[width=0.42\textwidth]{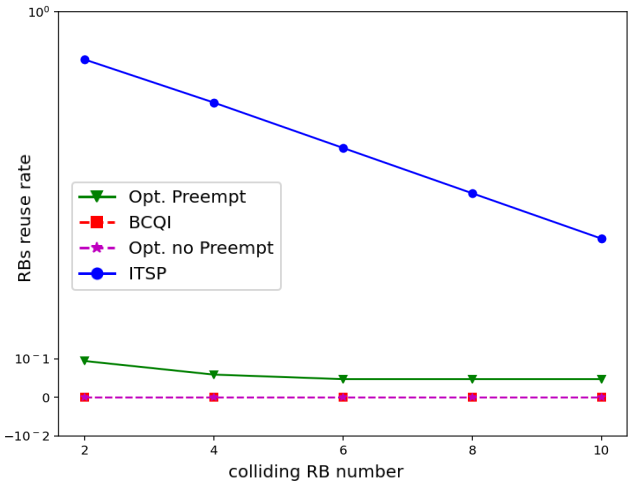} \\
(a) High load scenario ($\lambda_1 = 50, \lambda_2 = 10, D^{critic} = 5$) 
\vspace{5mm} \\
\includegraphics[width=0.42\textwidth]{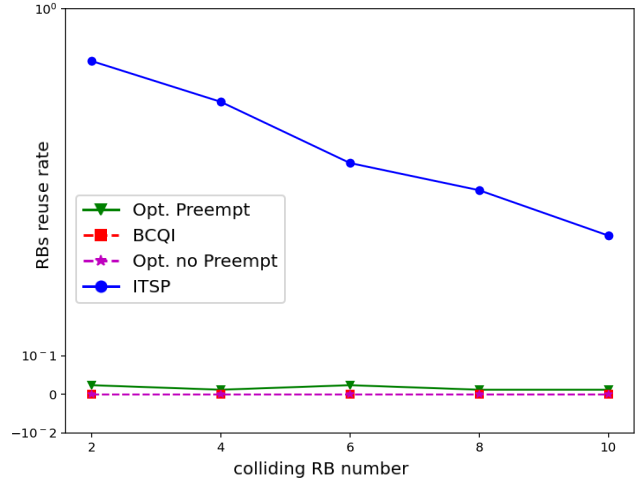} \\
(b) Low load scenario ($\lambda_1 = 50, \lambda_2 = 3, D^{critic} = 5$) 
\end{center}
{\caption{RBs reuse rate vs. Nbr. of colliding PRBs }
\label{fig14}}
\vspace{-3mm}
\end{figure}

Fig. \ref{fig14} shows the PRBs reuse rate as a function of the number of colliding PRBs. We can see that both BCQI and Opt. no preempt schemes always present a zero PRBs reuse rate. This is explained by the fact that these latter do not use preemption. Regarding ITSP, its use of preemption leads to a higher PRB reuse rate. It efficiently distributes the selected mini-slots needed for critical traffic across all available PRBs, while keeping the PRB allowance rate within limits. In contrast, the Opt. preempt solution does not exhibit this behavior.

\begin{figure}[t]%
\begin{center}
\includegraphics[width=0.42\textwidth]{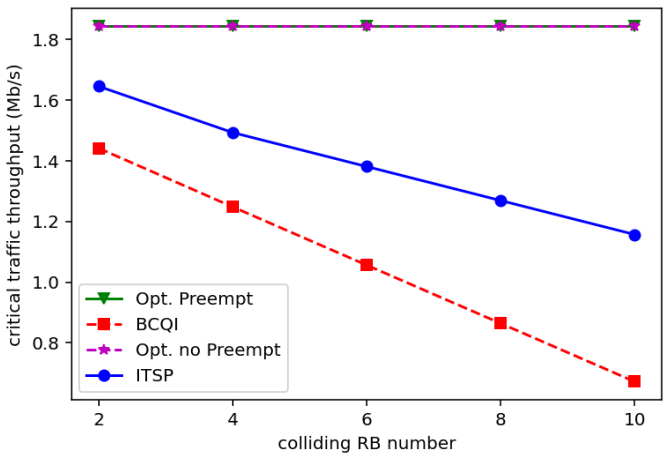} \\
High load scenario ($\lambda_1 = 50, \lambda_2 = 10, D^{critic} = 5$)
\end{center}
{\caption{Total Network Critical traffic throughput vs. Nbr. of colliding PRBs}
\label{fig22}}
\vspace{-3mm}
\end{figure}

Finally, Fig. \ref{fig22} shows the total network critical traffic throughput as a function of the number of colliding PRBs for all methods. As the behavior of the schedulers is slightly different, ITSP serves traffic on a per-arrival basis for each packet, while the Optimal solutions allocate resources based on the overall traffic received. Additionally, given the duration of the simulation, the ITSP scheme focuses only on serving packets nearing their expiration ($D^{critic}$), further contributing to the lower throughput compared to the optimal schedulers, which serve all traffic simultaneously.

\section{Conclusion}
This paper addresses the coexistence of GSM-R and FRMCS networks, focusing on FRMCS critical and performance traffic without impacting GSM-R traffic. We formulated the resource allocation problem as an ILP, considering GSM-R traffic and the optional use of preemption for prioritizing critical traffic. To solve this, we developed the ITSP scheduling algorithm, a CQI-aware approach that uses preemption to prioritize delay-sensitive critical traffic while optimizing performance traffic throughput. Simulations demonstrate that ITSP achieves high performance traffic throughput while maintaining acceptable critical traffic throughput.

\section*{Acknowledgment}
\begin{small}
This work was supported by the 5G-RACOM project, funded by the French Ministry of Economy, Finance and Industrial and Digital Sovereignty, in the innovation project program “Private 5G Networks for the Industry“.
\end{small}%


\end{document}